\documentclass[aps,twocolumn,showpacs,prc,citeonline,superscriptaddress,amsfonts,amsmath,amssymb]{revtex4-1}

\usepackage{graphicx}

\begin{document}

\title{Multitude of phases in correlated lattice fermion systems with spin-dependent disorder}

\author{J. Skolimowski}
\affiliation{Institute of Theoretical Physics,
Faculty of Physics,  University of Warsaw, ul.~Pasteura 5, PL-02-093 Warszawa, Poland}
\affiliation{Jo\v{z}ef Stefan Institute, Jamova 39, SI-1000 Ljubljana,Slovenia}

\author{D. Vollhardt}
\affiliation{Theoretical Physics III, Center for Electronic Correlations and Magnetism,
Institute of Physics, University of Augsburg, D-86135, Augsburg, Germany}

\author{K. Byczuk}
\affiliation{Institute of Theoretical Physics,
Faculty of Physics,  University of Warsaw, ul.~Pasteura 5, PL-02-093 Warszawa, Poland}

\date{\today}

\begin{abstract}

The magnetic phases induced by the interplay between disorder acting only on particles with a given spin projection (``spin-dependent disorder") and a local repulsive interaction is explored. To this end the magnetic ground state phase diagram of the Hubbard model at half-filling is computed within dynamical mean-field theory combined with the geometric average over disorder, which is able to describe Anderson localization.  Five distinct phases are identified: a ferromagnetically polarized metal, two types of  insulators, and two types of spin-selective localized phases. The latter four phases possess different long-range order of the spins.  The predicted phase diagram may be tested experimentally using cold fermions in optical lattices   subject to spin-dependent random potentials.

\end{abstract}

\pacs{ 71.10.Fd, 71.27.+a, 72.15.Rn, 67.85.Lm, 71.30.+h }

\maketitle
\section{Introduction}
Cold atoms in optical lattices provide an excellent experimental tool to explore  the interplay between interaction and disorder effects in quantum many-body systems \cite{Sanchez10}. Indeed, following  the seminal paper by Jaksch {\it et al}. \cite{jaksch98} ultracold atoms have been used to demonstrate a variety  of fundamental theoretical concepts, such as  the correlation-induced Mott transition \cite{greiner02,jordens08,Schneider08}, and the existence of a mobility edge in  non-interacting but disordered three-dimensional systems \cite{Semeghini15}.
The recent preparation of homogeneous Fermi gases of ultracold atoms in a \emph{uniform} potential \cite{Mukherjee17} will eventually make it possible to reproduce those solid state experiments since there is no longer a disturbance by a parabolic trapping potential. 
Furthermore, it is now even possible to   observe antiferromagnetic correlations \cite{Greif13, Hart15, parsons16, boll16, cheuk16, drewes16}. Together with a theoretically proposed new cooling method \cite{Paiva15}, this now allows  experiments with ultracold atoms to be performed at  temperatures  at which  antiferromagnetic order appears \cite{Mazurenko17}. Therefore, the problem of accessing the N\'eel temperature, below which the interacting particles are in an antiferromagnetic state, has been resolved \cite{Mazurenko17}.
These developments motivated us to extend our previous work on correlated lattice fermions with  spin-dependent disorder \cite{Skolimowski15,Makuch13} to the case with  antiferromagnetic long-range order (AF-LRO).  

The aim of this paper is to compute and discuss the magnetic ground state phase diagram of the Anderson-Hubbard model for spin $1/2$ fermions on a bipartite lattice at half-filling in the presence of spin-dependent disorder. Here "spin-dependent disorder"  means that disorder, i.e., randomly distributed local potentials, acts only on fermions with one particular spin direction \cite{nanguneri12,Makuch13,Skolimowski15}. 
Our previous studies showed that spin-dependent disorder strongly destabilizes the metallic phase. This is due to the breaking of the spin symmetry and, thus, the blocking of spin-flip processes which are responsible for quasiparticle formation. Moreover, a spin-selective localized phase was predicted \cite{Skolimowski15}. In this phase the particles with spin direction sensitive to randomness are localized, whereas particles in the opposite spin channel remain itinerant. The earlier studies of correlated lattice fermions with spin-dependent disorder \cite{nanguneri12,Makuch13,Skolimowski15} focused on paramagnetic phases,  i.e., phases without spontaneous long-range order (LRO). Thus  the question remained whether, and to what extent, AF-LRO of the fermions will change these results. In particular, two questions arise: \\ 
1) How does the possible existence of AF-LRO modify the paramagnetic ground state phase diagram of  interacting fermions with spin-dependent  disorder \cite{Skolimowski15}? \\
2) How does spin-dependent disorder change the antiferromagnetic ground state phase diagram obtained earlier \cite{Byczuk09} for interacting, disordered  fermions where the  disorder acts equally on both spin directions? \\
The present investigation provides answers to these questions. In particular, we show that now two spin-selective localized phases with LRO exists.  One such phase extends to arbitrarily strong disorder, and the system remains metallic in one of the spin-subsystem, in contrast to the case studied earlier \cite{Byczuk09}.  We also identify two different Mott insulating phases characterized  by ferrimagnetic spin order and ferromagnetic 
spin-density wave order, respectively.

In the following we will solve the Anderson-Hubbard model using the dynamical mean-field theory (DMFT) with geometric average over disorder. This non-perturbative approach is sensitive to  Anderson localization \cite{Anderson58} even on a one-particle level \cite{Dobrosavljevic03,dobro,mah} and treats disorder and interactions within a unified theoretical framework \cite{Byczuk05,dobro09,Byczuk11}. Replacing the arithmetic average employed in the coherent potential approximation \cite{Janis92} by the geometric average over the disorder  corresponds to the calculation  of the \emph{typical} local density of states (LDOS). Indeed, other DMFT studies of the Anderson-Hubbard model \cite{Dobrosavljevic97,Semmler11} have shown that the probability distribution function (PDF) of the LDOS approaches a log-normal distribution. For this PDF the geometric mean of random variables gives the most probable, i.e. typical,  value. Extended numerical investigations \cite{schubert10} and experimental studies \cite{richardella10}  provided evidence  that the log-normal distribution of the LDOS is actually an immanent feature of fermions close to Anderson localization.  The DMFT with geometric average has already been successfully employed to describe the  metal-insulator transition (MIT) at $T=0$ in a variety of interacting  models such as the Hubbard model \cite{Byczuk05, Byczuk09}, the Falicov-Kimball model \cite{byczuk05a, Souza07}, or a charge-transfer model \cite{Aguiar2014}, in the presence of disorder. It has also been used to examine the MIT in the paramagnetic,  disordered Hubbard model at finite temperatures \cite{Braganca15}.

\section{Model and method}
The Anderson-Hubbard model at half filling on a bipartite lattice with spin-dependent local disorder is described by the Hamiltonian 
\begin{eqnarray}
\label{AH}
H =  \sum_{ <i,j>\: \sigma} t_{ij} a^\dagger_{i\sigma}a_{j\sigma} + \sum_{i\sigma}\epsilon_{i\sigma} n_{i\sigma} \nonumber \\
+ U\sum_i \left( n_{i\uparrow}-\frac{1}{2}\right) \left(n_{i\downarrow}-\frac{1}{2}\right) ,
\end{eqnarray}
where $a_{i\sigma}$ ($a^\dagger _{i\sigma}$) is the fermionic annihilation (creation) operator of an electron at site $i$ and spin projection $\sigma =\pm 1/2=(\uparrow,\downarrow)$, $n_{i\sigma}=a^\dagger_{i\sigma}a_{i\sigma}$ is the particle number operator, and $U$ is the on-site repulsion. The hopping amplitude $t_{ij}$ is non-zero only between  nearest-neighbor sites $i$ and $j$. Due to this property the lattice is composed of two interpenetrating sublattices $s=\{A,B\}$. 

The local spin-dependent potentials $\epsilon_{i\sigma}$ are uncorrelated random variables drawn from a PDF  $\mathcal{P}_{\sigma}(x)$. Similar to our previous studies \cite{Makuch13, Skolimowski15}, the spin-dependent disorder is modeled by a box-shaped PDF  given by
\begin{equation}
\mathcal{P}_{\sigma}(x)=\frac{1+2\sigma}{2 \Delta}\Theta\left(\frac{\Delta}{2}-|x|\right),
\label{PDF}
\end{equation}
where  $\Theta (y)$ is  the Heaviside step function and $\Delta$ is the strength of the disorder. This means that the particles with spin up propagate on a lattice with randomly distributed on-site potentials, whereas the spin down particles move on an energetically uniform lattice. The PDF is the same on both sublattices, hence there is no dependence on the index $s$ in Eq.~(\ref{PDF}). Note that for the symmetric box-shaped PDF particle-hole symmetry holds. 

In order to include AF-LRO within DMFT  one has to treat the two sublattices separately. Hence for each sublattice $s$ the Hamiltonian (\ref{AH}) is mapped onto an ensemble of single-impurity Anderson models
\begin{eqnarray}
H^{s}_{\rm SIAM} = \sum_{\sigma} \epsilon_{\sigma} n_{\sigma} + U n_{\uparrow} n_{\downarrow} \nonumber \\
+\sum_{{\bf k}\sigma} (V_{{\bf k}\sigma, s} a^{\dagger}_{\sigma} c_{{\bf k} \sigma} + V_{{\bf k}\sigma, s }^*  c_{{\bf k} \sigma}^{\dagger} a_{\sigma}) + \sum_{{\bf k}\sigma} E_{{\bf k}\sigma,s} c_{{\bf k} \sigma}^{\dagger} c_{{\bf k} \sigma}
\label{AIM}
\end{eqnarray}
with random, spin-dependent on-site energies $\epsilon_{\sigma}$ drawn from the same PDF as in Eq. (\ref{PDF}). The last two  terms, describing the dispersion and the coupling of the  fermions of the bath to the impurity, are sublattice dependent.  The bath states are represented by a hybridization function 
\begin{equation}
\eta_{\sigma,s}(\omega)=\sum_{{\bf k}}\frac{|V_{{\bf k}\sigma, s}|^2}{\omega- E_{{\bf k}\sigma,s}},
\end{equation} which in the DMFT is determined self-consistently in the following way: For each $\epsilon_{\sigma}$ and sublattice $s$ we calculate the impurity Green function $G_{\sigma,s}(\omega, \epsilon_{\sigma})$ by solving the Hamiltonian (\ref{AIM}), and then determine the LDOS  via
\begin{equation}
\rho_{\sigma,s}(\omega,\epsilon_{\sigma})=-\frac{1}{\pi} \rm{Im} G_{\sigma,s}(\omega, \epsilon_{\sigma}) .
\end{equation}
 Next we find the geometrically averaged LDOS 
\begin{equation}
\rho_{\sigma,s}(\omega) = e^{ \langle \rm{ln} \rho_{\sigma,s} ( \omega,\epsilon_{\sigma} ) \rangle },
\end{equation}
 where $ \langle Q \rangle = \int d \epsilon \mathcal{P} _{\sigma}(\epsilon) Q(\epsilon)$ denotes the arithmetic average of $ Q(\epsilon) $.
 We note that the geometrically averaged LDOS, determined at finite disorder,  is not normalized to unity because it takes into account only extended states with  continuous spectrum. Therefore it is identically zero when all states are localized.  The localized states, having a dense point-like spectrum, are not taken into account here. 
The real part of the averaged local Green function is determined by  the Hilbert transform 
\begin{equation}
G_{\sigma, s} (\omega) = \int d \omega ' \frac{\rho_{\sigma,s} (\omega ')}{\omega - \omega '}.
\end{equation}
 The local self-energy $\Sigma _{\sigma, s} (\omega) $ is then obtained from the $\bf k$-integrated Dyson equation 
\begin{equation}
 \Sigma _{\sigma, s}(\omega) = \omega - \eta_{\sigma, s}(\omega) -  \frac{1}{G_{\sigma, s}(\omega)}.
\end{equation}
 The set of equations is closed by the Hilbert transform, which on a bipartite lattice is given by 
\begin{equation}
G_{\sigma,s}(\omega) = \int d \xi \frac{N_0(\xi) }{  \omega - \Sigma _{\sigma, s} (\omega ) - \frac{\xi^2}{\omega-\Sigma_{\sigma,\bar{s}}(\omega)} },
\end{equation}
 where $N_0(\xi)$ is the noninteracting density of states (DOS), and $\bar{s}$ denotes the sublattice opposite to $s$. These equations are solved iteratively until  self-consistency is reached. 

In our calculations the noninteracting DOS has the form $N_0(\xi) = 2 \sqrt{D^2-\xi^2}/\pi D^2$, where $W=2D$ is the bandwidth, and $W=1$ sets the energy unit. For this DOS the Hilbert transform can be obtained analytically, such that the local Green functions are related to the hybridization functions  by $ \eta_{\sigma,s}(\omega)=D^2 G_{\sigma,\bar{s}}(\omega)/4\;\;$ \cite{georges96}. 

The results were obtained by iteratively solving the DMFT equations at zero temperature using the numerical renormalization group (NRG) \cite{wilson,bulla2008}. For this part the open source NRG Ljubljana code \cite{nrgljubljana} was used.

\section{Phase diagram}

\begin{figure}[tbp]
	\begin{center}
		\includegraphics[clip,width=0.49\textwidth]{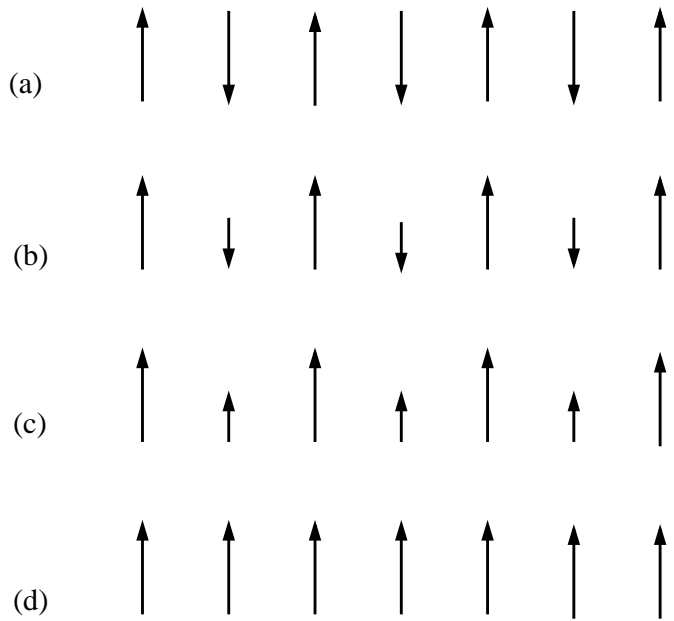}
		\caption{Panels (a)-(d) show different spin patterns found in the Anderson-Hubbard model with spin dependent disorder:  (a) antiferromagnet, (b) ferrimagnet, (c) ferromagnetic spin-density wave, and (d) ferromagnet.
}
		\label{magnetic_phases}
	\end{center}
\end{figure}

The ground state of the Anderson-Hubbard model Eq. (\ref{AH}) with spin-dependent disorder on a bipartite lattice is determined by three factors: the strengths of the disorder and of the local repulsion, respectively,  and the possible existence of AF-LRO.  They lead to the emergence of four different types of spin ordering, which are depicted schematically in panels (a)-(d) in Fig.~ \ref{magnetic_phases}. 
The  first pattern, (a), is the usual N\'eel AF-LRO, where the averaged spins are  of equal length, but are oriented anti-parallel  on neighbouring sites. 
The second pattern, (b), is a ferrimagnet, where the averaged spins on all sites are directed anti-parallel  but their  lengths on neighbouring sites differ. 
The  third pattern, (c),  is a ferromagnetic spin-density wave (SDW), where on neighbouring sites the  averaged spins are oriented parallel but have  different lengths. 
The last type of spin ordering, (d),    is the usual ferromagnet,  which is characterized by the same length and direction of  the averaged spins on every lattice site. 

To  fully characterize  the observed phases four different quantities are computed: \\ 
1) the geometrically averaged LDOS $\rho_{\sigma, s} (\omega)$ for spin $\sigma$ and sublattice $s$,\\ 
2) the  local magnetization $m_s$, i.e. the averaged value of the spin,  defined on a site belonging to sublattice~$s$ 
\begin{equation}
\label{ms}
m_s = \frac{n_{\uparrow,s} -n_{\downarrow,s}}{2},
\end{equation}
where $n_{\sigma, s}=\int_{-\infty} ^0 \rho_{\sigma, s} (\omega) d\omega$ is the density of particles with spin $\sigma$ on   sublattice $s$ (here the energy scale is chosen such that $\omega =0 $ corresponds to the Fermi energy),
\\ 3) the ferrimagnetic order parameter, i.e.,  the difference between the local magnetizations on neighbouring sites 
\begin{equation}
m_{\rm Ferri}= \frac{|m_A-m_B|}{2}\Theta\left(-\frac{m_A}{m_B}\right),
\label{mstg}
\end{equation}
and 
\\4)  the ferromagnetic spin-density wave order parameter, i.e.,  the difference between the parallel local magnetizations on neighbouring sites 
\begin{equation}
m_{\rm SDW} = |m_A - m_B|{\large \Theta}\left(\frac{m_A}{m_B}\right).
\end{equation}
The order parameters and the magnitude of the local magnetization, $|m_s|$, change between zero and $1/2$.

With these quantities one may identify  the following five different phases:\\
i) {\em ferromagnetic metal} (FM) if  $\rho_{\sigma, s} (0)\neq 0$ and  $m_A=m_B \neq 0$ (and consequently $m_{\rm Ferri} =0$ and $m_{\rm SDW} =0$),\\
ii) {\em ferrimagnetic insulator of type I} (Ins-I) if  $\rho_{\sigma, s} (0) = 0$, $m_{\rm Feri} \neq 0$, and $m_{\rm SDW}=0$, \\ 
iii) {\em ferromagnetic SDW insulator of type II} (Ins-II) if  $\rho_{\sigma, s} (0) = 0$, $m_{\rm Ferri} = 0$, and  $m_{\rm SDW}  \neq 0$, \\
iv)  {\em ferrimagnetic spin-selective localized phase of type I} (SSLP-I) if $\rho_{\uparrow s} (0) \neq 0$,  $\rho_{\downarrow s} (0) = 0$,  $m_{\rm Ferri} \neq 0$, and $m_{\rm SDW} = 0$\\
v)  {\em ferromagnetic SDW spin-selective localized phase of type II}  (SSLP-II) if $\rho_{\uparrow s} (0) = 0$ and $\rho_{\downarrow s} (0) \neq 0$ when $m_{\rm Ferri} = 0$, and $m_{\rm SDW} \neq 0$.\\
Within DMFT, a dynamical but local approximation, there are no other phases on a bipartite lattice. 

\begin{figure}[tbp]
	\begin{center}
		\includegraphics[clip,width=0.49\textwidth]{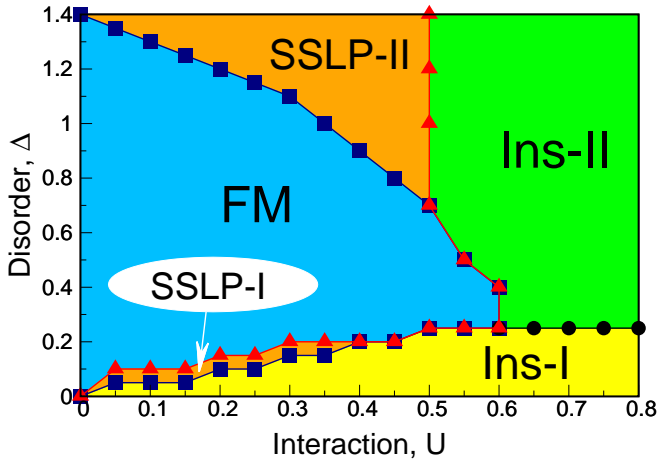}
		\caption{(Color online) The magnetic  ground state phase diagram of the Anderson-Hubbard model on a bipartite lattice at half filling  with spin-dependent disorder determined by DMFT with the geometrically averaged local density of states (LDOS). Five phases are distinguished: FM (Ferromagnetic metal), Ins-I  (insulator of type I), Ins-II (insulator of type II),  SSLP-I (spin-selective localized phase of type I), and SSLP-II  (spin-selective localized phase of type II).   At the point $\Delta=0$, $U=0$ the system is a paramagnet (Fermi gas) while on the horizontal line ($\Delta=0$, $U>0$) it is a N\'eel antiferromagnet. For details see text.}
		\label{phase_diagram}
	\end{center}
\end{figure}

The magnetic ground state  phase diagram of the Anderson-Hubbard model with spin-dependent disorder obtained in this way is shown in Fig.~\ref{phase_diagram} and is the main result of our investigation. It will now be discussed in detail.

\subsection{Ferrimagnetic insulator of type I (Ins-I)}
  
\begin{figure}[tbp]
\begin{flushleft}
	\includegraphics[clip,width=0.49\textwidth]{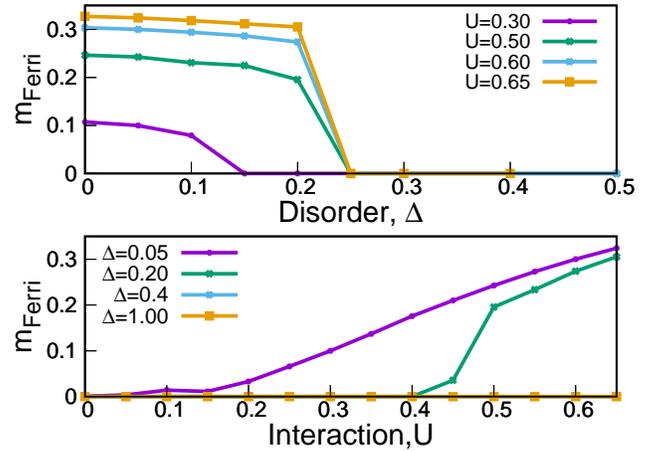}
		\caption{(Color online)  Ferrimagnetic order parameter $m_{\rm Ferri}$ as  a  function  of  disorder  strength  $\Delta$ for  different interaction strengths $U$ (upper panel), and  as a function of interaction strength $U$ for various
disorder strengths $\Delta$ (lower panel).}
		\label{magnetization1}
	\end{flushleft}
\end{figure}

This ferrimagnetic insulating phase is characterized by AF-LRO where the magnetization alternates in sign and magnitude on neighbouring sites, 
and  the  geometrically averaged LDOS vanishes at the Fermi energy in both spin channels and on every lattice site. In the absence of disorder ($\Delta=0$) this phase becomes an AF  insulator (AFI), which is the only stable DMFT solution for the half-filled Hubbard model on a bipartite lattice at $T=0$ and any $U>0$, due to particle-hole and spin symmetries. The ferrimagnetic insulator exists only at weak disorder $\Delta$, i.e., in the regimes with $\Delta \lesssim U/3$ for $U\lesssim 0.6$ and $\Delta \lesssim 0.25$ for $U\gtrsim 0.6$.  Upon increasing the disorder the ferrimagnetic order parameter  $m_{\rm Ferri}$ goes to zero in both interaction regimes as is seen in the upper panel of Fig.~\ref{magnetization1}. The vanishing of $m_{\rm Ferri}$ signals a possible transition to a different spin pattern. At the same time, by turning up $U$ the ferrimagnetic order parameter increases and the system tends toward the saturated N\'eel antiferromagnet with $m_{\rm Ferri}=1/2$, as shown by the curves for $\Delta=0.05$ and $0.2$ in the lower panel of Fig.~\ref{magnetization1}.

\begin{figure}[tbp]
	\begin{center}
\includegraphics[clip,width=0.49\textwidth]{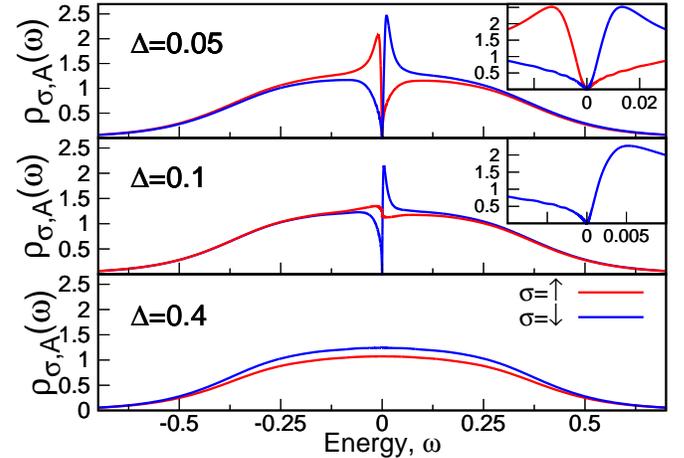}
	\caption{(Color online) Geometrically averaged LDOS on sublattice $A$ for different values of the disorder strength $\Delta$ at interaction strength $U=0.2$. Results are shown for  Ins-I (upper panel), SSLP-I (middle panel), and FM (lower panel). Insets show zoomed regions around the Fermi level.}
	\label{DOS_U_0_20}
	\end{center}
\end{figure}

The geometrically averaged LDOS \cite{comment_ldos} for the ferrimagnetic insulator for $U\lesssim 0.6$ exhibits a narrow spectral gap and a pronounced asymmetric peak (cf. the upper panel in Fig. \ref{DOS_U_0_20}) similar to the $\Delta=0$  case \cite{Pruschke03}. 
Due to the staggered (N\'eel) spin order the lattice unit cell is doubled in the ferrimagnetic insulator. The existence of LRO in this phase is caused by the interaction, while the ferrimagnetic modulation of the local magnetization is an effect of the spin-dependent disorder. This type of disorder reduces the band-width of fermions with spin up  in extended states.  Since the interaction is weak here, the LDOS for spin down particles is not strongly modified by spin-dependent disorder and almost remains the same as in the $\Delta=0$ case. This altogether leads to staggered, but different in magnitude, local magnetizations $|m_A|\neq |m_B|$ and ferrimagnetic order.

\subsection{Ferrimagnetic spin-selective localized phase of type I (SSLP-I)}

 In the weakly interacting regime with $U\lesssim 0.6$ the ferrimagnetic insulator undergoes  a transition to the ferrimagnetic spin-selective localized phase upon increase of the disorder $\Delta$.  This phase  is shown as a narrow orange area in  the phase diagram in Fig.~\ref{phase_diagram}. The SSLP-I phase  is characterized by small values of $m_{\rm Ferri}$ as is seen in  Fig.~\ref{magnetization1}  and a non-vanishing  geometrically averaged LDOS at the Fermi energy in only one spin channel, as is illustrated in the middle panel of Fig.~\ref{DOS_U_0_20}. This means that particles with spin up, which  are directly influenced by the disorder, are now in the metallic phase because the disorder redistributes their spectral weight, thereby closing the gap for these fermions. By contrast, the spin down particles remain in the insulating state. In other words, particles with spin down, which are not directly affected  by the spin-dependent disorder, sustain the  gap in the geometrically averaged LDOS due to the existence of AF-LRO. 
Although the gap is closed in $\rho_\uparrow(\omega)$ the asymmetry of the geometrically averaged LDOS in the two spin channels remains, giving rise to different local magnetizations $|m_A|\neq |m_B|$ and a finite value of  $m_{\rm Ferri}$.

\subsection{Ferromagnetic metal (FM)}

\begin{figure}[tbp]
\begin{center}
	\includegraphics[clip,width=0.49\textwidth]{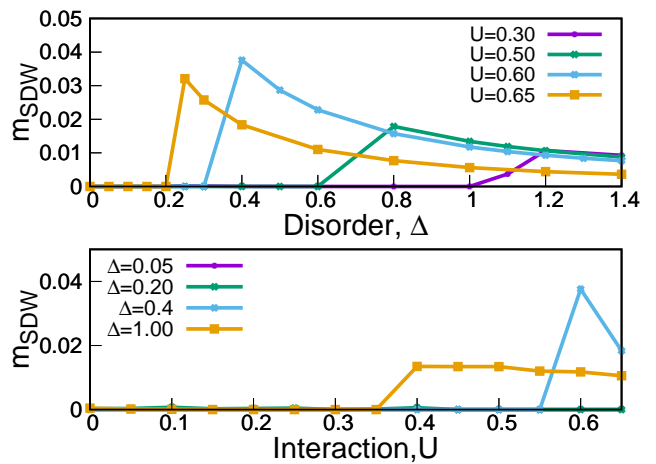}
		\caption{(Color online) Ferromagnetic SDW order parameter as  a  function  of  disorder  strength  $\Delta$ for  different interaction strengths U (upper panel), and as a function of interaction strength U for various disorder strengths $\Delta$ (lower panel).}
		\label{magnetization2}
	\end{center}
\end{figure}

\begin{figure}[tbp]
	\begin{center}
		\includegraphics[clip,width=0.49\textwidth]{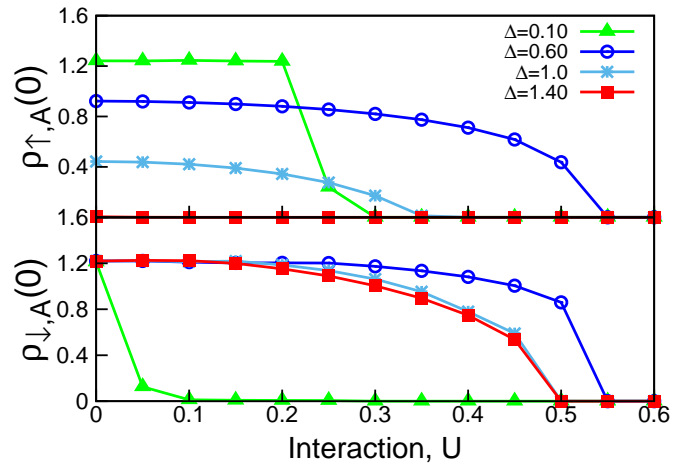}
		\caption{(Color online) Geometrically averaged LDOS on sublattice $A$ at the Fermi level as a function of interaction  $U$ for different values of the disorder $\Delta$.  Spin-up particles: upper panel,  spin-down particles: lower panel.}
		\label{dos_zero_D}
	\end{center}
\end{figure}

 Upon increasing the disorder strength further the FM becomes stable with finite and equal local magnetizations $m_s$  on both sublattices. This phase  does not possess a spontaneous LRO since the uniform spin polarization is driven solely by the spin-dependent disorder. This phase is characterized by  $m_{\rm Ferri}=0$  and $m_{\rm SDW}=0$ (c.f., Figs.~\ref{magnetization1},~and~\ref{magnetization2}) whereas the geometrically averaged LDOS at the Fermi energy for both spin species becomes non-zero, as is seen in Fig.~\ref{dos_zero_D}.  

In this ferromagnetic metallic phase spin-dependent disorder plays a dominant role. Indeed, it closes the gaps in the geometrically averaged LDOS in the two spin channels and leads to an equal distribution of spectral weights below and above the Fermi level, as is shown in the lower panel of Fig.~\ref{DOS_U_0_20}. 
Since disorder reduces the spectral weight of spin up particles, c.f., the lower panel of Fig. \ref{DOS_U_0_20},  the system becomes spin polarized with $|m_A|= |m_B|\neq 0$. The  interaction (repulsion) thereby plays a minor role here: namely, it mediates the influence of the disorder also to the spin down particles. As a result, both spin up and spin down particles are distributed uniformly on both sublattices, which leads to the absence of  LRO  ($m_{\rm Ferri}=0 $ and $m_{\rm SDW}=0 $).

\subsection{Ferromagnetic SDW spin-selective localized phase of type II (SSLP-II)}

When the disorder strength is increased even further  a transition from the FM phase to the ferromagnetic SDW spin-selective localized phase takes place at $U\lesssim 0.5$. In the SSLP-II the spin-up particles, which are influenced directly by spin-dependent disorder, have a vanishing  geometrically averaged  LDOS at Fermi level on both sublattices: $\rho_{\uparrow,s}(0) = 0$, i.e., they are in an insulating state. By contrast,  the spin-down particles have $\rho_{\downarrow,s}(0)\neq 0$ on both sublattices, and hence  are  metallic. At the same  time the ferromagnetic SDW is stabilized, with $m_{\rm SDW}$ being relatively small, as  seen in Fig.~\ref{magnetization2}. This SSLP-II is a magnetic counterpart to the spin-selective localized phase found in the paramagnetic ground state phase diagram in Ref.~\onlinecite{Skolimowski15}.

\begin{figure}[tbp]
	\begin{center}
		\includegraphics[clip,width=0.49\textwidth]{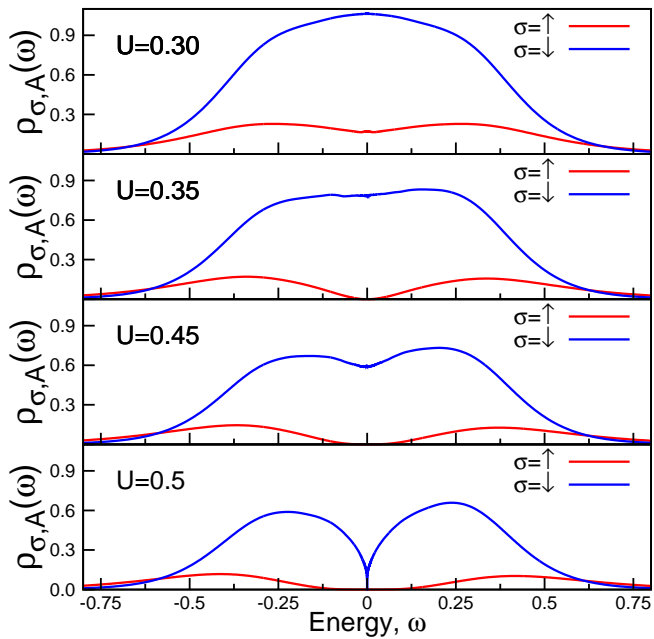}
		\caption{(Color online) Geometrically averaged LDOS on sublattice $A$ for different values of the interaction $U$ at disorder strength $\Delta=1$. Top panel: FM; second panel from the top: border between  FM and SSLP-II; third panel from the top: SSL-II; bottom  panel: border between SSLP-II and  AFI.}
		\label{DOS_D_1_00}
	\end{center}
\end{figure}

The origin of this phase can be explained as follows: strong spin-dependent disorder renormalizes the geometrically averaged LDOS  of the spin-up particles and  opens a gap  at the Fermi level due to the disorder-driven localization transition, thus forming two narrow, continuous subbands. The changes of the geometrically averaged LDOS, and the spin-selective opening of the gap when $U$ is increased at fixed $\Delta=1$, is shown in the panels of Fig.~\ref{DOS_D_1_00}. The properties of SSLP-II can be  effectively understood within a Falicov- Kimball model \cite{Falicov69,Freericks03}  where spinless fermions on a lattice interact with immobile particles. Indeed, the spin up particles  are localized due to spin-dependent disorder and act as scatterers for the spin down particles due to the Hubbard interaction $U$. In contrast to the paramagnetic case studied in Ref.~\onlinecite{Skolimowski15} here the parallel oriented local magnetic moments have different values on different sublattices, yielding a small but finite ferromagnetic SDW order parameter $m_{\rm SDW}$. Spin-selective localization together with ferromagnetic SDW  LRO implies the absence of spin up quasiparticles at the Fermi level.
They are, however, present for spin-down particles as seen for $U=0.35$ and $0.45$ in  Fig.~\ref{DOS_D_1_00}. We also see in Fig.~\ref{DOS_D_1_00} that  Hubbard subbands at higher energies are formed in the geometrically averaged LDOS for both spins.  

\begin{figure}[tbp]
	\begin{center}
\includegraphics[clip,width=0.4\textwidth]{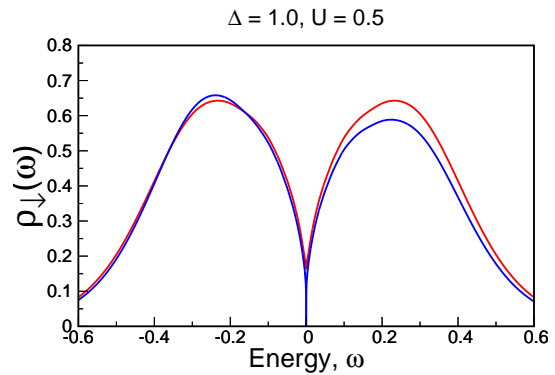}
		\caption{(Color online) Comparison of the geometrically averaged LDOS on sublattice $A$ for $\Delta=1$ and $U=0.5$ with (red curve) and without (blue curve) AF-LRO.}
		\label{AFvsPM}
	\end{center}
\end{figure}

At the interaction $U\approx 0.5$ the SSLP-II turns into a ferromagnetic SDW  insulator (Ins-II) as is seen in Fig.~\ref{phase_diagram}, and the geometrically averaged LDOS shows a gap for both spin particles, cf. the lower panel of Fig.~\ref{DOS_D_1_00}. Similar to the paramagnetic case discussed in Ref.~\onlinecite{Skolimowski15}, the critical interaction $U_c$, at which the transition from SSLP-II to Ins-II takes place, is independent of the disorder strength $\Delta$ as  shown in  Fig.~\ref{phase_diagram}. We conclude that the transition from SSLP-II to Ins-II is of the Falicov-Kimball type because the geometrically averaged LDOS for spin-down particles splits in a similar way as in the case of binary-alloy disorder, where particles with spin-up provide  localized scattering centers \cite{Freericks03}. The slight reduction of the critical interaction from $U_c\approx 0.55$ for the paramagnetic ground state discussed in Ref.~\onlinecite{Skolimowski15}, to $U_c\approx 0.5$ in the case studied here, is caused by the presence of the ferromagnetic SDW LRO which naturally tends to form a gap and subbands, as in the Slater theory of AF \cite{Pruschke03}. Indeed, this is seen in Fig.~\ref{AFvsPM}, where we compare the geometrically averaged LDOS at $\Delta=1$ and $U=0.5$ obtained from  the uniform and bipartite lattice DMFT solutions, respectively.  The ferromagnetic SDW LRO leads to an asymmetric transfer of the spectral weight away from the Fermi level and an opening of the gap  at smaller $U$ as compared with the uniform case. We also note that for the disorder values considered here $m_{\rm SDW}$ is almost independent of $\Delta$, which we can see in  Fig.~\ref{magnetization2}. Therefore the $U_c(\Delta)$ line is vertical in the phase diagram Fig.~\ref{phase_diagram}.

\subsection{Ferromagnetic SDW insulator (Ins-II)}

\begin{figure}[tbp]
	\begin{center}
\includegraphics[clip,width=0.5\textwidth]{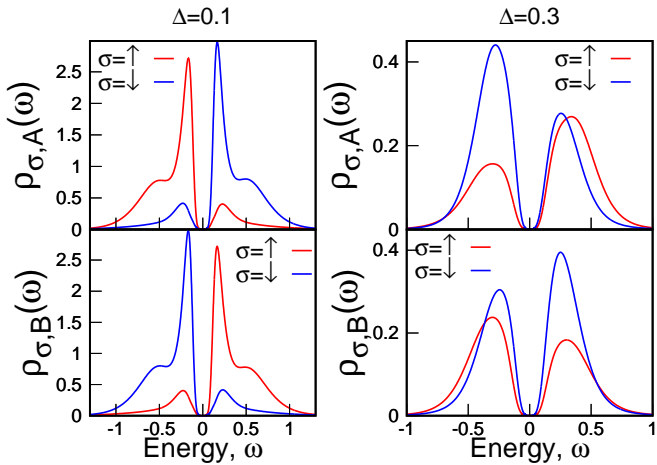}
		\caption{(Color online) Comparison of geometrically averaged LDOS on sublattice A and sublattice B for Ins-I (left-hand side) and Ins-II (right-hand side) at the same $U=0.7$.}
		\label{Ferro_vs_Ferri}
	\end{center}
\end{figure}

 Once the interaction strength exceeds $U \approx 0.6$ for $\Delta \gtrsim 0.2$, the system makes a transition from the ferrimagnetic insulator (Ins-I) with $m_{\rm Ferri}\neq 0$ to the ferromagnetic SDW insulator (Ins-II) with $m_{\rm SDW}\neq 0$. Due to the spin-dependent bandwidth renormalization caused by the localization transition, the density of spin up particles is reduced when the disorder strength is increased. Above  $\Delta \approx 0.2$ the reduction is so strong that, although there is an interaction driven tendency toward formation of AF-LRO in the system, the spin-dependent disorder together with the localization transition of spin up particles drives the system into the ferromagnetic SDW order, c.f the left and right panel of Fig.~\ref{Ferro_vs_Ferri}. Thereby this type of disorder reduces the role of virtual exchange processes responsible for AF magnetic LRO at large $U$ \cite{spalek07}. On the other hand, unequal but parallel local magnetization on two sublattices keeps the system in the insulating phase with $m_{\rm SDW}\neq 0$. In the case of disorder acting equally on both spin directions, this part of the  phase diagram would be an antiferromagnetic insulator and the interaction would play a dominant role, generating AF-LRO \cite{Byczuk09}. 

\section{Conclusions and outlook}

In this paper we extended the investigation of correlated lattice fermions with spin-dependent disorder by including the effects of antiferromagnetic long-range order on the ground state properties.  Apart from a ferromagnetic metal, we found two insulating and two spin-selective localized phases. The two insulating phases differ in the pattern of their spin ordering, one having parallel and the other  staggered orientation on the sublattices. In the two spin-selective localized phases particles with either up or down spin orientation are localized. The phase diagram and properties of the ground state were discussed in detail. It is surprising that such a simple model can lead to a phase diagram with such a multitude of different phases. Obviously, an extension of the present study to finite temperatures is called for. 

Spin-dependent disorder can be realized experimentally by focusing light beams with different polarization, after having been scattered from a diffusive plate, on an optical lattice  \cite{Mandel03,McKay10,Makuch13,Skolimowski15}. This, together with recently developed methods for cooling \cite{Paiva15} and  detecting antiferromagnetic correlations \cite{Greif13, Hart15, parsons16, boll16, cheuk16, drewes16,Mazurenko17}, will make it possible to explore correlated lattice fermions in the presence of spin-dependent disorder  experimentally and to test our predictions.

\begin{acknowledgments}

Support by the Deutsche Forschungsgemeinschaft through TRR 80 is gratefully acknowledged.

\end{acknowledgments}



\begin{thebibliography}{}
\bibitem{comment_ldos} Because of the symmetry $\rho_{\sigma A}(\omega) = \rho_{\bar{\sigma}B}(-\omega) $ only LDOS on  sublattice $A$ are presented.
\end{thebibliography}
\end{document}